\documentclass[aps,superscriptaddress,nofootinbib,twocolumn,truncate=10]{revtex4-2}
\usepackage{bm}
\usepackage{lmodern}

\usepackage[utf8]{inputenc}
\usepackage{amsmath}
\usepackage{amssymb}
\usepackage{upgreek}
\usepackage{graphicx}
\usepackage{epstopdf}
\usepackage[dvipsnames]{xcolor}
\usepackage{enumerate}
\usepackage{gensymb}
\usepackage{ulem}
\usepackage[T1]{fontenc}
\newcommand{\stkout}[1]{\ifmmode\text{\sout{\ensuremath{#1}}}\else\sout{#1}\fi}
\usepackage[english]{babel}
\usepackage{braket}
\usepackage{float}
\usepackage{stmaryrd}
\usepackage{mathtools}
\usepackage[misc]{ifsym}
\usepackage{siunitx}

\newcommand{\AI}[1]{\textcolor{black}{#1}}

\newcommand\blankfootnote[1]{%
  \let\thefootnote\relax\footnotetext{#1}%
  \let\thefootnote\svthefootnote%
}
\newenvironment{myfont}{\fontfamily{ptm}\selectfont}{\par}
\DeclareTextFontCommand{\textmyfont}{\myfont}

\usepackage{hyperref}
\usepackage{lipsum}

\begin{document}

\title{Operating semiconductor qubits without individual barrier gates} 

\author{Alexander S. Ivlev}
\thanks{These authors contributed equally}
\affiliation{QuTech and Kavli Institute of Nanoscience, Delft University of Technology, PO Box 5046, 2600 GA Delft, The Netherlands}
\author{Damien R. Crielaard}
\thanks{These authors contributed equally}
\affiliation{QuTech and Kavli Institute of Nanoscience, Delft University of Technology, PO Box 5046, 2600 GA Delft, The Netherlands}
\author{Marcel Meyer}
\affiliation{QuTech and Kavli Institute of Nanoscience, Delft University of Technology, PO Box 5046, 2600 GA Delft, The Netherlands}
\author{William I. L. Lawrie}
\affiliation{QuTech and Kavli Institute of Nanoscience, Delft University of Technology, PO Box 5046, 2600 GA Delft, The Netherlands}
\author{Nico W. Hendrickx}
\affiliation{QuTech and Kavli Institute of Nanoscience, Delft University of Technology, PO Box 5046, 2600 GA Delft, The Netherlands}
\author{Amir Sammak}
\affiliation{QuTech and Netherlands Organisation for Applied Scientific Research (TNO), Delft, The Netherlands}
\author{\AI{Yuta Matsumoto}}
\affiliation{QuTech and Kavli Institute of Nanoscience, Delft University of Technology, PO Box 5046, 2600 GA Delft, The Netherlands}
\author{\AI{Lieven M. K. Vandersypen}}
\affiliation{QuTech and Kavli Institute of Nanoscience, Delft University of Technology, PO Box 5046, 2600 GA Delft, The Netherlands}
\author{Giordano Scappucci}
\affiliation{QuTech and Kavli Institute of Nanoscience, Delft University of Technology, PO Box 5046, 2600 GA Delft, The Netherlands}
\author{Corentin Déprez}
\affiliation{QuTech and Kavli Institute of Nanoscience, Delft University of Technology, PO Box 5046, 2600 GA Delft, The Netherlands}

\author{Menno Veldhorst}
\email{m.veldhorst@tudelft.nl}
\affiliation{QuTech and Kavli Institute of Nanoscience, Delft University of Technology, PO Box 5046, 2600 GA Delft, The Netherlands}

\date{\today}

\begin{abstract}
Semiconductor spin qubits have emerged as a promising platform for quantum computing, following a significant improvement in their control fidelities over recent years. Increasing the qubit count remains challenging, beginning with the fabrication of small features and complex fanouts. A particular challenge has been formed by the need for individual barrier gates to control the exchange interaction between adjacent spin qubits. Here, we propose a method to vary two-qubit interactions without applying pulses on individual barrier gates while also remaining insensitive to detuning noise in first order. \AI{Experimentally we find that} changing plunger gate voltages over \SI{300}{mV} \AI{can} tune the exchange energy $J$ from $\SI{100}{kHz}$ to $\SI{60}{MHz}$. This allows us to perform two-qubit operations without changing the barrier gate voltage. Based on these findings we conceptualize a spin qubit architecture without individual barrier gates, simplifying the fabrication while maintaining the control necessary for universal quantum computation. 
\end{abstract}

\maketitle

A large qubit count is essential for achieving fault-tolerant quantum computing and quantum advantage. While various platforms are scaling to tens and even hundreds of qubits~\cite{kim_2023_Nature_IBMEagle,decrossComputationalPowerRandom2025,bluvstein_2024_Nature_LogicalQuantumProcessor,acharyaQuantumErrorCorrection2025}, a similar increase in the number of qubits with gate-defined quantum dots has remained challenging~\cite{hendrickx_2021_Nature_FourqubitGermaniumQuantum,philips_2022_Nature_UniversalControlSixqubit,george12SpinQubitArraysFabricated2025}. Even though their operational fidelities are competitive~\cite{Noiri_2022_Nature_Threshold,Xue_2022_Nature_SurfaceCodeThreshold,mills_2022_ScienceAdv_TwoqubitSiliconQuantum,Wang_2024_Science_OperatingSemiconductorQuantum}, and their scaling prospects are promising~\cite{zwerver_2022_NatureElectronics_QubitsMadeAdvanced,neyens_2024_Nature_ProbingSingleElectrons300mm,steinacker_2024_arxiv_300MmFoundry}, the limited size of the current semiconductor spin systems so far prevents the platform to host meaningful quantum error correction experiments or large analog simulations~\cite{bloch_2012_NaturePhys_QuantumSimulationsUltracold,scholl_2021_Nature_QuantumSimulation2D,bluvstein_2024_Nature_LogicalQuantumProcessor,acharyaQuantumErrorCorrection2025,karamlou_2024_Nature_ProbingEntanglement2D}. The small array sizes moreover limit the ability to gather detailed statistics to improve the device designs, further slowing down the progress of the platform. \\

A key challenge in increasing the quantum dot array size is set by the small feature size of the quantum dot structures. The fabrication of these finer structures generally suffers in terms of resolution, uniformity and yield, in particular when many need to be placed near one another. The introduction of hole spin devices in germanium has \AI{shown that increased feature sizes}~\cite{lodari_PRB_LightEffectiveHole2019} \AI{can lead} to the rapid growth of the quantum dot count, even in two dimensions~\cite{hendrickx_naturecomm_GatecontrolledQuantumDots2018,vanriggelen_APL_TwodimensionalArraySinglehole2021,Wang_2024_Science_OperatingSemiconductorQuantum,Borsoi_2024_NatureNano_SharedControl16}. \AI{Increasing} the feature sizes may thus be a key step toward\AI{s} even larger quantum dot arrays\AI{, both for holes in germanium and for platforms where the charge carriers have a large effective mass, such as electrons and holes in silicon}. It is therefore worth to reconsider the critical components of quantum dot devices necessary for future experiments. Current state-of-the-art devices~ \cite{mills_2022_ScienceAdv_TwoqubitSiliconQuantum,philips_2022_Nature_UniversalControlSixqubit,Wang_2024_Science_OperatingSemiconductorQuantum,takeda_npj_RapidSingleshotParity2024,george12SpinQubitArraysFabricated2025} generally consist of individual plunger gates separated by dedicated barrier gates. Barrier gates in particular have a smaller footprint making them more difficult to contact through vias and to fabricate. They are typically also more numerous as individual barrier gates are patterned in between each neighbouring quantum dot pair. Thus, the use of individual barrier gates leads to a significant overhead, hindering the fabrication of larger arrays.
\\

In this work, we take steps towards removing individual barrier gates, which are commonly used to tune the exchange interaction necessary for two-qubit gates. We introduce an alternative method to perform exchange operations without applying barrier gate pulses while staying at the charge symmetry point\AI{, where sensitivity to charge noise is reduced}. Instead, the exchange interaction can be altered by simultaneously increasing the chemical potential of two neighbouring quantum dots. To increase the range of control of this method, two germanium quantum dots are isolated from the reservoir, allowing us to change their chemical potential by ten times their charging energy. Notably, we see no significant decay in spin coherence over this large voltage range. In this regime, we show that our method allows to vary the exchange interaction \AI{by more than} two orders of magnitude, enabling the implementation of a controlled phase operation without pulsing on the barrier gate. Based on these findings, we propose a spin qubit architecture that does not rely on individual barrier gates. 

\section{Methods to tune the exchange coupling}

\begin{figure*}
\includegraphics[width=\textwidth]{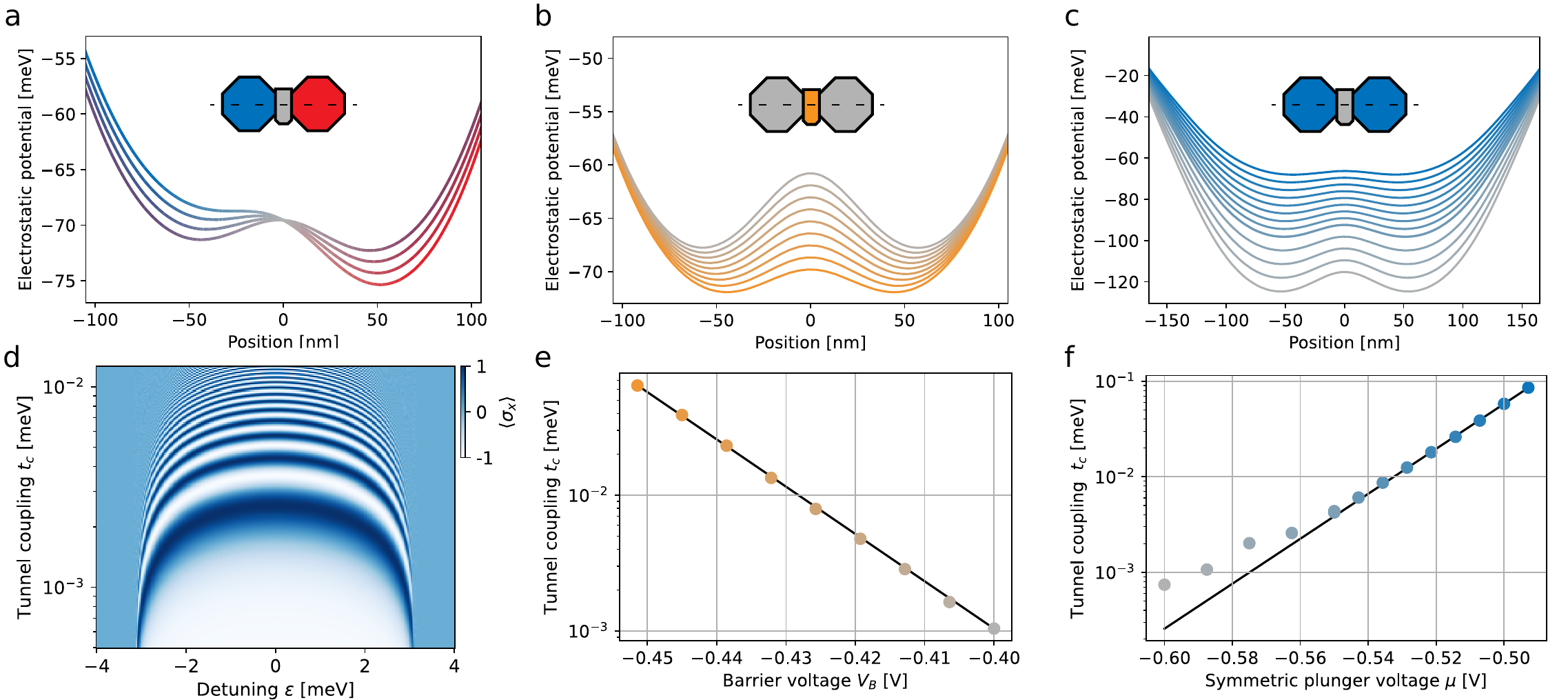}
\caption{\textbf{Methods for manipulating the exchange coupling between spin qubits.} \AI{\textbf{(a)} Simulated electrostatic potential in the quantum well when the detuning $\varepsilon$ is varied to control the exchange coupling $J$. The inset depicts the device used during the simulation, as well as the dashed linecut along which the potential is plotted. The brighter colours correspond to larger detuning and thus larger exchange values. Details on the simulations performed can be found in Supplementary Material~I~\cite{supp_reference}). We note that to first order the tunnel coupling $t_c$ does not change as $\varepsilon$ is varied. \textbf{(b)} Simulated electrostatic potential when the barrier gate voltage $V_B$ is used to control $J$ through the tunnel coupling $t_c$. The line-colour corresponds to the voltage on the marker in \textbf{e} with the same colour. The plunger gate voltages are fixed at $\mu=-\SI{0.5}{mV}$. \textbf{(c)} Simulated potential when the symmetric plunger voltage $\mu$ is used to control $J$ through $t_c$, at $V_B=-\SI{0.45}{mV}$. The line-colour corresponds to the voltage on the marker \textbf{f}  with the same colour. \textbf{(d)} Simulated fingerprint plot~\cite{reed_PRL_ReducedSensitivityChargeNoiseSymmetricGates} showing tunability of the exchange energy $J$ when the tunnel-coupling $t_c$ and detuning $\varepsilon$ are varied. The ensemble-average projection $\langle \sigma_x \rangle$ of a superposition state is depicted under the realistic assumptions of detuning noise and charging energy~\cite{hendrickxSweetspotOperationGermanium2024,stehouwerExploitingEpitaxialStrained2025}. Details can be found in Supplementary Material~I~\cite{supp_reference}). \textbf{(e)} Simulated $t_c$ as function of $V_B$. The solid line is an exponential fit to the simulated data. \textbf{(f)} $t_c$ is simulated as function of $\mu$. Contrary to the barrier-based variation, the $t_c$ does not have a pure exponential dependence with $\mu$, which we hypothesise is related to the larger deformation of the confinement potential.}}
\label{fig:SchematicsDevice_And_Principle}
\end{figure*}

To eliminate barrier-based control of the exchange interaction, it is essential to understand the parameters that contribute to this exchange. The exchange energy $J\approx \frac{4Ut_c^2}{U^2-\varepsilon^2}$ between spins generally depends on the detuning $\varepsilon$ between neighbouring dots as well as the interdot tunnel coupling $t_c$~\cite{burkard_2023_RevModPhys_SemiconductorSpinQubits}. Here $\varepsilon=0$ corresponds to the charge symmetric point and \AI{$U\gg t_c$ }is the on-site Coulomb interaction. In early experiments, the exchange interaction was manipulated using the detuning $\varepsilon$, as depicted in \AI{the simulated electrostatic potential of} Figure~\ref{fig:SchematicsDevice_And_Principle}a. This allows to control the exchange interaction in simple devices and it enabled pioneering work in the field~\cite{petta_2005_Science_CoherentManipulationCoupled,veldhorst_2015_Nature_TwoqubitLogicGate,watson_2018_Nature_ProgrammableTwoqubitQuantum}. However, due to the strong influence that detuning has on the exchange coupling at the interaction point, this method suffers from large charge noise susceptibility\AI{~\cite{PhysRevLett.110.146804}}. 
\\

Noise resilience can be improved by staying at the symmetric charge sweet spot $\varepsilon=0$, where $J$ is \AI{to} first-order insensitive to noise on $\varepsilon$, and instead changing the tunnel coupling $t_c$~(see \cite{reed_PRL_ReducedSensitivityChargeNoiseSymmetricGates,martins_2016_PRL_NoiseSuppressionUsingSymmetricExchangeGates} \AI{and the simulated fingerprint plot in Figure~\ref{fig:SchematicsDevice_And_Principle}d)}. \AI{The tunnel coupling is} commonly \AI{controlled} using a dedicated barrier gate as depicted in Figure~\ref{fig:SchematicsDevice_And_Principle}b\AI{,e} and it allows for high-fidelity two-qubit operations~\cite{Noiri_2022_Nature_Threshold,Xue_2022_Nature_SurfaceCodeThreshold,mills_2022_ScienceAdv_TwoqubitSiliconQuantum,Wang_2024_Science_OperatingSemiconductorQuantum}. However, a third method can be envisioned, in which the tunnel coupling \AI{is} controlled without changing the barrier gate voltage. Indeed by simultaneously varying the chemical potential of both quantum dots \AI{$\mu$} as depicted in Figure~\ref{fig:SchematicsDevice_And_Principle}\AI{c}, one also changes the tunnel barrier and with it the exchange coupling\AI{,} using only plunger gates. Crucially, \AI{$t_c$ and thus $J$ can be controlled this way} while staying at the charge sweet spot $\varepsilon=0$. \AI{We confirm this tunability of $t_c$ in Figure~\ref{fig:SchematicsDevice_And_Principle}f with a Schr\"odinger-Poisson simulation with the Luttinger-Kohn Hamiltonian in QTCAD~\cite{QTCADAssistedDesign} (for details see Supplementary Material~I~\cite{supp_reference}).} We will refer to this \AI{method of tuning $t_c$ using $\mu$} as symmetric barrier-free pulsing.
\section{Operating an isolated spin qubit pair}

\begin{figure*}
\includegraphics[width=\textwidth]{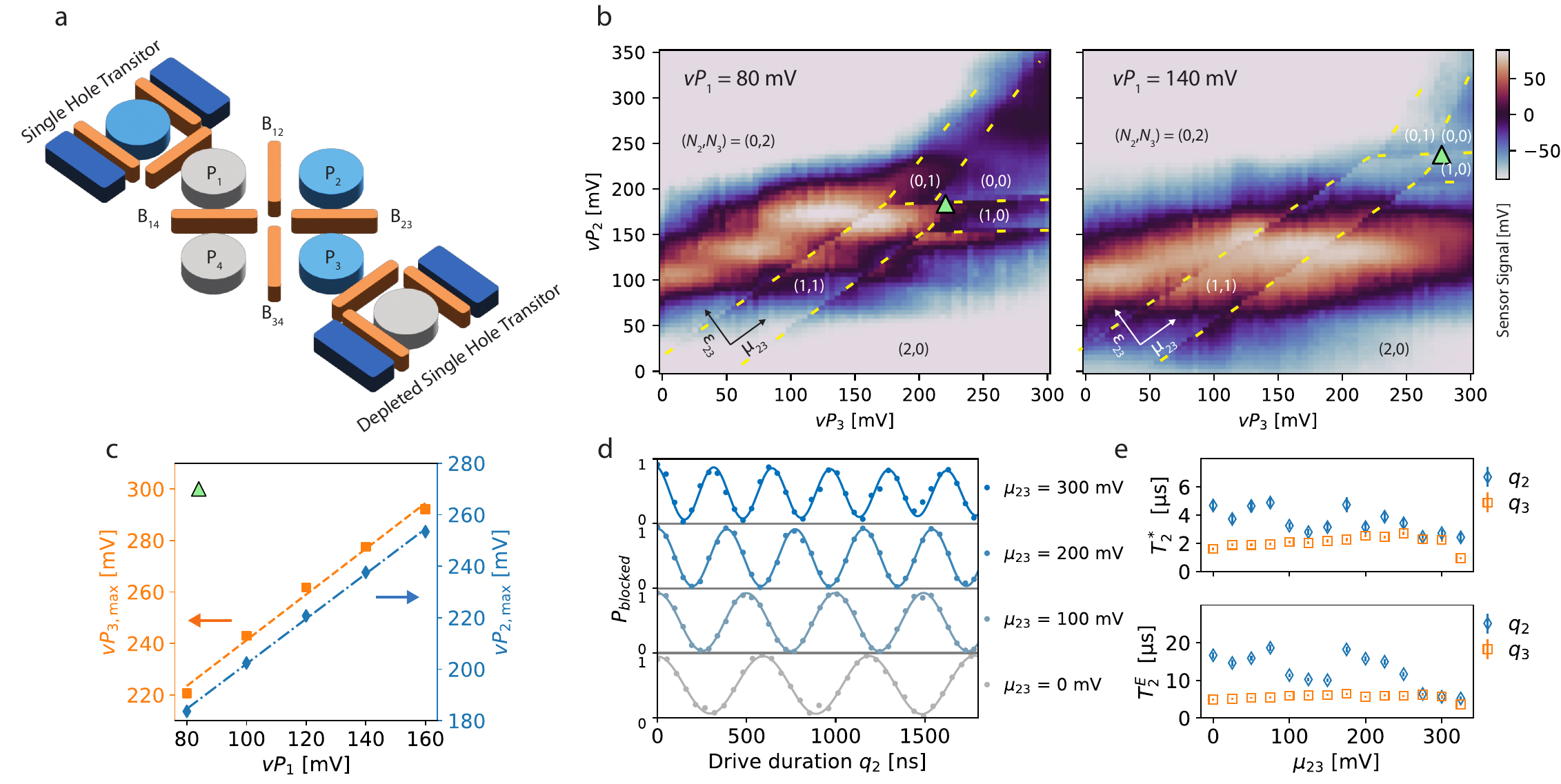}
\caption{\textbf{Operating qubits in the isolated regime.} \textbf{(a)} Schematic representation of the device used in this experiment. The blue colour denotes the occupied quantum dots under P\textsubscript{2} and P\textsubscript{3}, and the top single-hole transistor acts as a reservoir. The grey quantum dots under P\textsubscript{1}, P\textsubscript{4} and the bottom single-hole transistor remain depleted. The dark blue elements are the ohmic contacts and the various barrier gates are given in orange. The external magnetic field is fixed at $B_0=\SI{50}{mT}$ approximately in-plane, parallel to the line connecting P\textsubscript{2} and P\textsubscript{3}. \textbf{(b)} Charge stability diagrams of the isolated two-hole regime using virtual plunger gates $\text{\textmyfont{\textit{v}}}P_{2}$ and $\text{\textmyfont{\textit{v}}}P_{3}$ for different values of $\text{\textmyfont{\textit{v}}}P_{1}$. The detuning $\varepsilon_{23}$ and chemical potential $\mu_{23}$ axes are indicated. The charge occupation $(N_{\text{2}},N_{\text{3}})$ of quantum dots 2 and 3 is denoted in each charge region. \AI{The dashed lines serve as guide to the eye indicating charge transitions}. For each quantum dot just two transitions are visible. The absence of additional charge transitions confirms the system is in the isolated regime. The maximal voltages $\text{\textmyfont{\textit{v}}}P_{\text{2,max}}$ and $\text{\textmyfont{\textit{v}}}P_{\text{3,max}}$ for which the quantum dots remain in a (1,1) occupation are marked by the green triangle. These voltages increase linearly with $\text{\textmyfont{\textit{v}}}P_{\text{1}}$ as seen in subfigure \textbf{(c)}, where the dashed lines indicate a linear fit of the maximal voltages as a function of $\text{\textmyfont{\textit{v}}}P_{1}$. \textbf{(d)} Rabi oscillations induced by EDSR driving of $q_{2}$ at $\epsilon_{23}=0$ for several values of $\mu_{23}$. The solid line gives a fit to an exponentially decaying sinusoidal with visibilities up to $0.89\pm0.01$. \textbf{(e)} The Ramsey and Hahn-echo coherence times  $T_2^*$ and $T_2^E$ as a function of $\mu_{23}$, fitted using an exponentially decaying sine. The error bars give the standard deviation of the fitted coherence.}
\label{fig:Verify_Operational_region}
\end{figure*}

To study the symmetric barrier-free pulse scheme and compare it to the existing schemes, we use a Ge/SiGe qubit device with dedicated barrier gates between each plunger gate. This $2\times2$ quantum dot device, schematised in Figure~\ref{fig:Verify_Operational_region}a, and the underlying heterostructure have been introduced previously in respectively~\cite{hendrickx_2021_Nature_FourqubitGermaniumQuantum} and~\cite{lodari_MatQuantTech_LowPercolationDensity2021}. \AI{Details on the fabrication of the heterostructure and gate-stack can be found in the Supplementary Material~II~\cite{supp_reference})}. The device has two single-hole transistors positioned diagonally across each other. During this experiment, the single-hole transistor next to plunger gate P\textsubscript{1} functions as both a charge sensor and a hole reservoir. Underneath plunger gates P\textsubscript{2} and P\textsubscript{3} we initialise a pair of hole spin qubits $q_2$ and $q_3$ in the $\ket{\downarrow \downarrow}$ state. \AI{While hole spins are used in this experiment, the underlying physics is understood to be the same for electrons, and the experiments can therefore be interpreted in the context of gate-defined quantum dot devices in general.} Details on the readout and initialisation can be found in the Methods section. To control the exchange interaction $J_{23}$ between the spins we either change the detuning $\varepsilon_{23}=\frac{1}{2}(\text{\textmyfont{\textit{v}}}P_2-\text{\textmyfont{\textit{v}}}P_3)$, the virtualised barrier voltage $\text{\textmyfont{\textit{v}}}B_{23}$ or the chemical potential on each quantum dot $\mu_{23}=\frac{1}{2}(\text{\textmyfont{\textit{v}}}P_2+\text{\textmyfont{\textit{v}}}P_3)$. All gate voltages are virtualised to compensate for the cross-talk on the chemical potential of adjacent quantum dots~\cite{hensgens_2017_Nature_QuantumSimulationFermi}. \AI{We stress that this cross-talk is compensated with the corresponding plunger gate voltages and not the voltages on barrier gates B\textsubscript{ij}. This means that when $\mu_{23}$ or $\varepsilon_{23}$ are altered the voltage applied on the barrier gates remains constant, and the pulse can be considered barrier-free as intended.}\\

To avoid any spurious interactions and unwanted redistribution of charges, the quantum dots under P\textsubscript{1}, under P\textsubscript{4} and the bottom single-hole transistor are depleted during operation~\AI{\cite{eeninkTunableCouplingIsolation2019,tanttuAssessmentErrorsHighfidelity2024,bertrandQuantumManipulationTwoElectron2015}}. This prevents any charges from loading onto or escaping from the quantum dots under P\textsubscript{2} and P\textsubscript{3} during operation through energetically favourable states on the neighbouring sites. As a result, after the initial loading of two charges (see Methods for loading sequence), no additional charges are loaded even after decreasing the chemical potential $\mu_{23}$ significantly. This is confirmed by the extended interdot transitions in the charge stability diagrams of Figure~\ref{fig:Verify_Operational_region}b. Moreover, the voltages at which charges are unloaded are dictated by the chemical potential of the neighbouring quantum dots. Indeed we see in Figures~\ref{fig:Verify_Operational_region}b-c, that the range over which the (1,1) charge state is maintained is linearly dependent on the voltage $\text{\textmyfont{\textit{v}}}P_{1}$. Increasing the voltage on neighbouring plunger gates therefore enlarges the range of $\mu_{23}$ that is applicable for the symmetric barrier-free operations. This isolation further ensures the absence of neighbouring spins, preventing any spurious exchange interaction while varying the chemical potential. In the remainder of the presented data, the virtualised voltage on plunger gate 1 is set to $\text{\textmyfont{\textit{v}}}P_{1}=\SI{240}{mV}$. At this voltage, $\mu_{23}$ could be varied by roughly $\SI{350}{mV}$ while staying in the (1,1) charge state. This range corresponds to ten times the charging voltage of $\SI{35}{mV}$, as extracted from the charge stability diagrams. \AI{The DC voltages on the barrier gates B\textsubscript{12}, B\textsubscript{23}, B\textsubscript{34} are set to be approximately $\SI{-250}{mV}$, mimicking the constraints of a global barrier architecture that is introduced later in the manuscript.}\\

Given the accessible range of $\mu_{23}$, we first check that the spin qubits remain coherent and can be operated after applying large voltage pulses on the plunger gates. We perform standard qubit characterisation at different values of $\mu_{23}$ using Rabi, Ramsey and Hahn-echo experiments. Some examples of Rabi oscillations obtained via electric dipole spin resonance (EDSR) driving at different values of $\mu_{23}$ are depicted in Figure~\ref{fig:Verify_Operational_region}d. Above $\mu_{23}=\SI{325}{mV}$, Rabi oscillations started to degrade and EDSR driving was not possible. We hypothesise that this loss of EDSR signal at high $\mu_{23}$ results from a charge transition to the quantum dot underneath P\textsubscript{1}. Having demonstrated the ability to drive the qubits, we characterise their coherence at different $\mu_{23}$ in Figure~\ref{fig:Verify_Operational_region}e. We do not observe a pronounced trend of the Ramsey ($T_2^*$) or Hahn-echo ($T_2^E$) coherence as a function of $\mu_{23}$, apart from the coherence decrease at higher values which we attribute to an increased exchange coupling \AI{and with it a larger sensitivity to charge noise}. The fact that coherence is generally maintained over such a large range in $\mu_{23}$ is encouraging for the symmetric barrier-free operation introduced here, as it suggests that a large tunability of the tunnel-coupling can be achieved without undermining the individual qubit operation. 

\begin{figure*}
\includegraphics[width=0.8\textwidth]{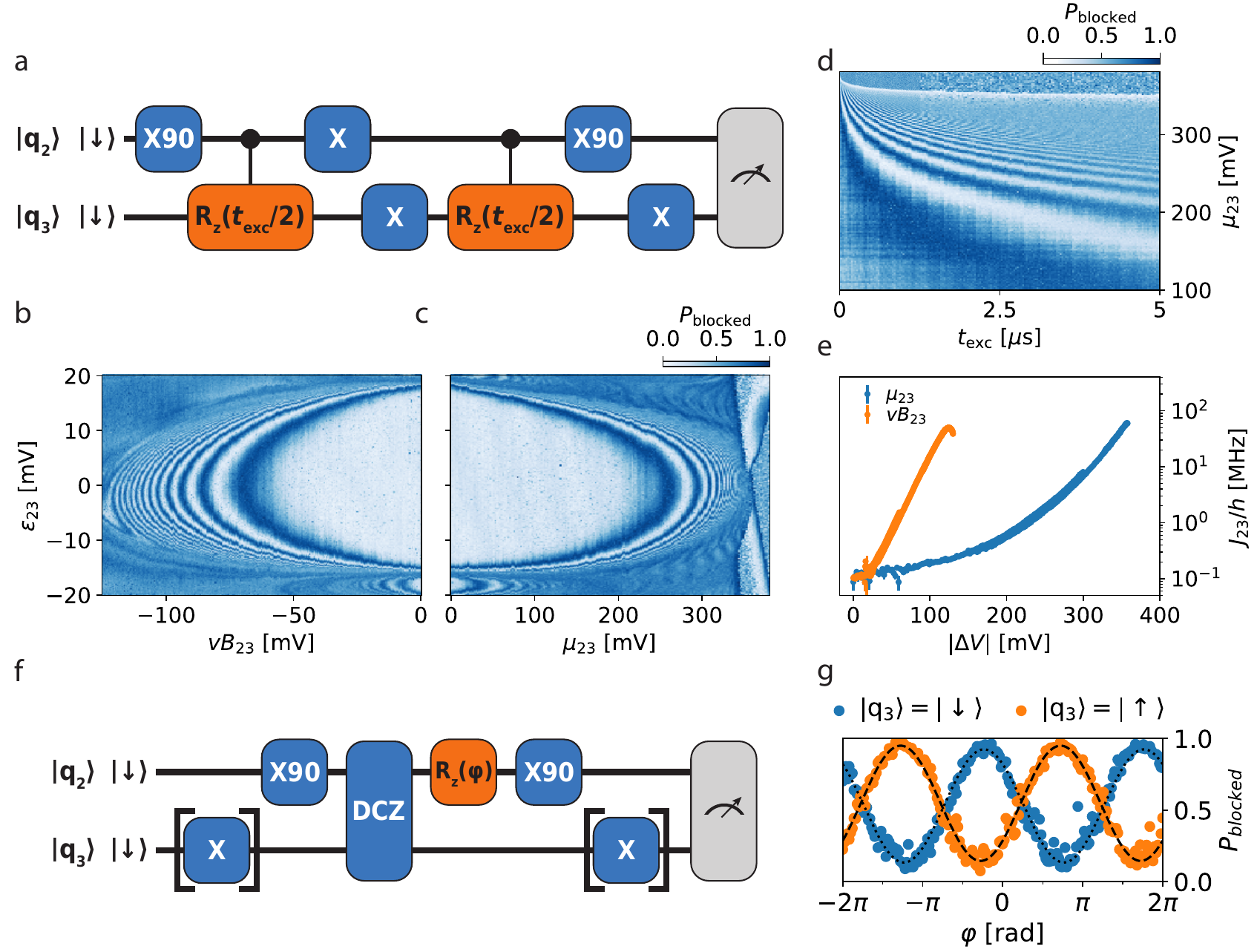}
\caption{\textbf{Controlling exchange interaction with and without barrier gate pulses.} \textbf{(a)} The gate sequence used to determine the exchange interaction $J$. $q_2$ is put into an equal spin superposition. The exchange is turned on during two controlled phase blocks, each lasting $t_{\text{exc}}/2$, separated by two decoupling pulses that cancel any \AI{single-qubit} phase accumulation during the exchange pulse. Final single-qubit rotations allow for readout of distinguishable states. The blocks highlighted in orange are varied throughout subfigures \textbf{b-e}. \textbf{(b-c)} Fingerprint plots corresponding to the circuit in \textbf{a}, where the exchange time is fixed at $t_{\text{exc}}=\SI{0.5}{\micro s}$ and the gate voltages are varied. The origin of both plots corresponds to the same voltage. $\textbf{(d)}$ Exchange oscillations as a function of the exchange time $t_{\text{exc}}$ at the symmetry point $\epsilon_{23}=\SI{0}{mV}$. \textbf{(e)} Exchange energy $J_{23}$ extracted from fitting the time-resolved oscillations in \textbf{d} and similar exchange oscillations induced by pulsing on the barrier gate. \textbf{(f)} The circuit used to demonstrate perform controlled-Z pulses. $q_2$ is initialised in an equal spin superposition and an optional spin-flip is performed on $q_3$. The decoupled controlled-Z (DCZ) pulse consists of two Tukey-window exchange pulses with a maximum height of $\mu_{23}=\SI{275}{mV}$, separated by a double spin-echo. To measure the phase of $q_2$ we perform a virtual rotation around the z-axis, with varying phase $\varphi$. If $q_3$ was initialised in $\ket{\uparrow}$, it is flipped back for readout (see Methods). \textbf{(g)} State evolution observed while implementing circuit \textbf{f}. The oscillations obtained for an initial $q_3=\ket{\uparrow}$ state are approximately $\pi$ shifted compared to those obtained for an initial $q_3=\ket{\downarrow}$ state, showing that the DCZ allows to perform a controlled-Z operation.}
\label{fig:Exchange_data}
\end{figure*}

\section{Controlling the exchange}
Now that the coherence of each individual qubit at varying $\mu_{23}$ is established, we can study the effect of $\mu_{23}$ on the exchange interaction between the two spins. We extract exchange oscillations using the sequence depicted in Figure \ref{fig:Exchange_data}a. Here the exchange interaction is turned on twice for a time $t_{\text{exc}}/2$ with a double decoupling pulse in between~\cite{russHighfidelityQuantumGates2018,watson_2018_Nature_ProgrammableTwoqubitQuantum}. This double decoupling pulse cancels out single-qubit phase accumulation induced by the exchange pulses, while the two-qubit conditional phase due to $ZZ$ interactions is still being accumulated. Moreover, the decoupled sequence results in a longer coherence time of the state allowing to probe smaller values of the exchange interaction $J_{23}$~\cite{wang_2024_arxiv_PursuingHighfidelityControl}. Using this technique we conduct a fingerprint scan~\cite{reed_PRL_ReducedSensitivityChargeNoiseSymmetricGates} as a function of the virtual barrier voltage $\text{\textmyfont{\textit{v}}}B_{\text{23}}$ and detuning $\varepsilon_{23}$ at a fixed total exchange time $t_{\text{exc}}=\SI{0.5}{\micro s}$. The resulting fingerprint plot can be found in Figure~\ref{fig:Exchange_data}b. When the barrier gate voltage is decreased, more fringes appear due to the increasing exchange energy $J_{23}$\AI{, as we'd expect from the simulated fingerprint in Figure~\ref{fig:SchematicsDevice_And_Principle}d}. We repeat this experiment, but varying $\mu_{23}$ instead of $\text{\textmyfont{\textit{v}}}B_{\text{23}}$ (Figure~\ref{fig:Exchange_data}c). Crucially, now $J_{23}$ increases when the voltage $\mu_{23}$ is increased, mirroring the dependence on $\text{\textmyfont{\textit{v}}}B_{\text{23}}$. The fingerprint also confirms that we remain at the charge symmetry point as $\mu_{23}$ is swept along $\varepsilon_{23}=0$. Similar to exchange pulses using dedicated barrier gates, it is therefore possible to keep low charge noise sensitivity while using only the plunger gates. We note that the feature around $\mu_{23}\approx\SI{350}{mV}$ coincides with the charge transition seen in the charge stability diagrams. Besides fingerprint plots, we further confirm the exchange tunability by studying the exchange splitting through EDSR measurements, which can be found in Supplementary Material~III~\cite{supp_reference}). Similar fingerprint plots are obtained for the quantum dot pair \AI{QD\textsubscript{1}-QD\textsubscript{4}} in Supplementary Material~IV~\cite{supp_reference}).\\

To study the dependence of the exchange energy in more detail we vary the exchange time $t_{\text{exc}}$ in the pulse sequence of Figure~\ref{fig:Exchange_data}a at $\epsilon_{23}=0$. The time-resolved oscillations as a function of $\mu_{23}$ are shown in Figure~\ref{fig:Exchange_data}d, with similar results for $\text{\textmyfont{\textit{v}}}B_{23}$. The frequency of these oscillations equals $J_{23}/2$, which we extract by fitting a decaying sinusoid. In Figure~\ref{fig:Exchange_data}e we plot the extracted dependence of $J_{23}$ on $\text{\textmyfont{\textit{v}}}B_{23}$ and $\mu_{23}$. Over a large range of the barrier gate voltage, we see the expected exponential dependence of the exchange energy which is fitted to obtain a tunability of $36.6\pm0.1\,\SI{}{mV/dec}$, comparable to state-of-the-art devices~\cite{george12SpinQubitArraysFabricated2025}. We understand the apparent non-monotonicity of $J_{23}$ at higher values of $\text{\textmyfont{\textit{v}}}B_{\text{23}}$ to result from the level-crossing of the singlet $\ket{S}$ and triplet $\ket{T_-}$ states (see Supplementary Material III~\cite{supp_reference})). Moreover, we observe that $J_{23}$ saturates at low values of $|\Delta V_B|$. Our method does not allow to extract lower exchange couplings considering the coherence times of our qubits. Looking at the symmetric barrier-free pulses, we see that changing $\mu_{23}$ allows us to tune $J_{23}$ by approximately a factor of 600, from 100~kHz to 60~MHz. The maximal exchange reachable is limited by the available range of $\mu_{23}$ as discussed earlier. We further note that the exchange interaction does not follow an exponential dependence on $\mu_{23}$. \AI{This is in line with the earlier simulation of tunnel coupling in Figure~\ref{fig:SchematicsDevice_And_Principle}e and we speculate that this results from a significant change in the confinement potential (see Figure~\ref{fig:SchematicsDevice_And_Principle}c)}. To achieve the same maximum increase in effective exchange coupling, $\mu_{23}$ needs to be pulsed almost 3 times as much as the barrier gate voltage. We hypothesise this is partly explained by the barrier gate not only changing the potential barrier height but also its width, as seen in Figure~\ref{fig:SchematicsDevice_And_Principle}b. Moreover, this difference could result from the higher lever arm of $\text{\textmyfont{\textit{v}}}B_{\text{23}}$ compared to that of the plunger gates, as the former is fabricated in a lower gate layer~\cite{hendrickx_2021_Nature_FourqubitGermaniumQuantum}. Given the limited swing of an AWG, increasing the lever arm of the plunger gates would in general allow to more effectively isolate the system and enlarge the range over which $\mu_{23}$ can be pulsed. In future devices, the plunger gates can be fabricated at a lower gate layer to gain these benefits. Already in the current device, however, the range of $\mu_{23}$ is large enough to access similar values of the exchange energy as with the dedicated barrier gate, and achieve on-off ratios comparable to the state-of-the-art devices~\cite{Xue_2022_Nature_SurfaceCodeThreshold,philips_2022_Nature_UniversalControlSixqubit,weinsteinUniversalLogicEncoded2023}. \\

\AI{In the current experiment, the quality of the exchange oscillation is higher for the barrier-based pulses with up to $Q_{B_{23}}=20\pm2$ compared to the symmetric barrier-free exchange oscillation $Q_{\mu_{23}}=10\pm2$, while we find that the QD\textsubscript{1}-QD\textsubscript{4} pair has $Q_{\mu_{14}}=18\pm4$ (see Supplementary Material~V~\cite{supp_reference}) for details). The variation in the barrier-free quality factor may come from the specific potential landscape and charge noise present.}\\

\begin{figure}
\includegraphics[width=1\columnwidth]{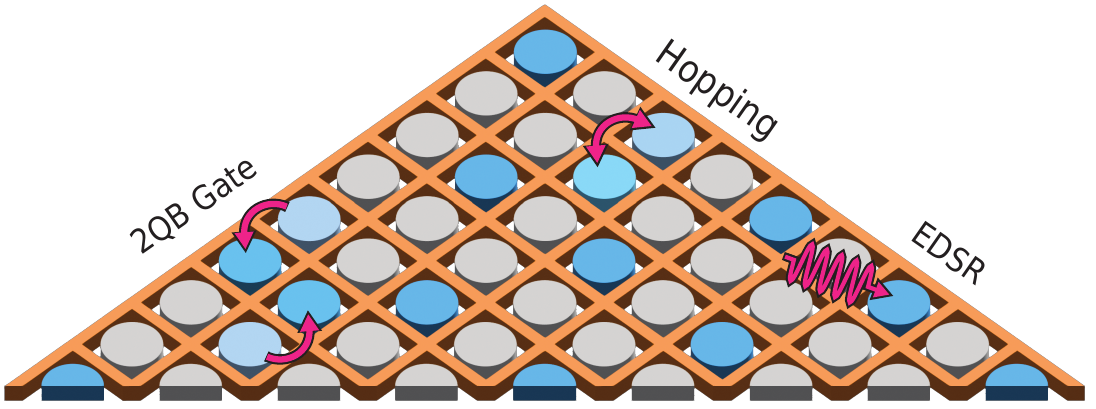}
\caption{\textbf{Merged barrier architecture and operation.} A clipped section of the proposed scalable quantum dot array architecture. The individual barrier gates are merged into a single barrier grid (orange). The quantum dots are sparsely occupied (blue). The empty quantum dots (grey) can be used to connect distant qubits or provide space for qubit operation. The exact choice of occupation generally depends on the desired functionality of the empty quantum dots. Charge sensors can be integrated within the array itself or placed on the periphery. Arrows indicate possible operations like EDSR, hopping-based single-qubit gates and two-qubit (2QB) gates through shuttling.}
\label{fig:CZ_and_Architecture}
\end{figure}

By controlling the exchange energy with symmetric barrier-free pulses it is possible to realise state-dependent interactions without pulsing on dedicated barrier gates. We use this to demonstrate a decoupled controlled-Z operation between $q_{\text{2}}$ and $q_{\text{3}}$ after preparing $q_{\text{2}}$ in an equal superposition (Figure~\ref{fig:Exchange_data}f). For each exchange pulse, we pulse to $\mu_{23}=\SI{275}{mV}$ using a Tukey window~\cite{Xue_2022_Nature_SurfaceCodeThreshold,Wang_2024_Science_OperatingSemiconductorQuantum,rimbach-russSimpleFrameworkSystematic2023} to enhance adiabaticity. After the two-qubit operation, we study the state-dependent phase accumulation, with the result depicted in Figure~\ref{fig:Exchange_data}g. After the decoupled exchange pulse, the superposition state of $q_{\text{2}}$ gains a relative phase of $(1.041\pm0.006)\pi$ when $q_{\text{3}}$ is flipped, as is expected from a controlled-Z pulse.\\

\section{A merged grid architecture}

Having shown the ability to perform entangling operations without using a dedicated barrier, we incorporate this technique into \AI{an} architecture design for large-scale semiconductor quantum processors. In this architecture, the numerous barrier gates are merged into a large two-dimensional barrier grid, as schematically depicted in Figure~\ref{fig:CZ_and_Architecture}. The plunger gates are still separately connected to allow for individual control over the chemical potential of each quantum dot. The array is generally sparsely occupied, with the spin qubits only brought together during two-qubit operations, \AI{initialisation and readout}. The sparse occupation enables the use of the symmetric barrier-free pulses investigated here without creating spurious interactions. \\

In this architecture, the global barrier grid helps to confine a spin to a single quantum dot \AI{and coarsely sets the tunnel coupling between adjacent quantum dots,} while control \AI{over each individual} plunger gate \AI{enables universal qubit operation.} Single-qubit gates can be performed through EDSR, or through spin-hopping~\cite{Wang_2024_Science_OperatingSemiconductorQuantum}. Two-qubit gates are executed by moving two spins together and performing the symmetric barrier-free operations \AI{demonstrated in this manuscript. We emphasise that the symmetric, detuning-insensitive aspect of this method allows the global barrier grid to be a viable alternative to individual barrier gate control}. Two-qubit \AI{gates} can be further tuned by varying the time that qubits spend adjacent to each other. \AI{Finally, while this manuscript focussed on controlling the exchange interaction, it is expected that initialisation and readout can be achieved without individual barrier control in a similar fashion (see the discussion in the Methods). Symmetric barrier-free pulses can again be used to tune the tunnel coupling to the values desired for high-fidelity initialisation and readout~\cite{takeda_npj_RapidSingleshotParity2024,nurizzoCompleteReadoutTwoElectron2023}. }\\

To have full control over each physical qubit, the number of electronic connections still grows linearly with the number of qubits in this merged barrier architecture. Yet, with the removal of individual barrier \AI{gates}, the distance between leads is increased. \AI{This increase of effective footprint simplifies the} local fanout significantly. Moreover, since plunger gates are typically larger than barrier gates, these remaining plunger gates are easier to contact using vias. These simplifications could allow spin-qubit array sizes to increase in two dimensions more easily, which is valuable for fault-tolerant computation~\cite{bluvstein_2024_Nature_LogicalQuantumProcessor,acharyaQuantumErrorCorrection2025}.\\

We further note that the sparse occupation required results in an additional overhead compared to a dense occupation. This is deemed acceptable given the functionalities a sparse architecture provides, like integrated shuttling lanes \cite{seidler_2022_npjQuant_ConveyormodeSingleelectronShuttling,zwerver_2023_PRX_ShuttlingElectronSpin,kunne_2024_NatureComm_SpinBusArchitectureScaling,vanriggelendoelman_2024_NatureCom_CoherentSpinQubit} for increased connectivity to support efficient error correction~\cite{bravyi_2024_Nature_HighthresholdLowoverheadFaulttolerant}, space for high-fidelity hopping-based gates~\cite{Wang_2024_Science_OperatingSemiconductorQuantum,unseldBasebandControlSingleelectron2024} and reduced cross-talk due to increased qubit spacing~\cite{johnTwodimensional10qubitArray2024}. Still, a dense occupation of the quantum dot array is conceivable in the context of analog quantum simulation experiments, where the global barrier grid allows to turn on all interactions between the neighbouring quantum dots simultaneously\AI{, albeit with some natural variability}. This architecture would then allow quantum dot experiments~\cite{vandiepen_2021_PRX_QuantumSimulationAntiferromagnetic,wang_2023_npjQuantInfo_ProbingResonatingValence} to more readily reach the sizes for analog simulation that have been achieved in atomic and superconducting systems~\cite{bloch_2012_NaturePhys_QuantumSimulationsUltracold,scholl_2021_Nature_QuantumSimulation2D,karamlou_2024_Nature_ProbingEntanglement2D}.\\

In conclusion, we have investigated the effect of symmetrically pulsing on both plunger gates of a double quantum dot. We have seen that in an isolated regime, these pulses do not significantly limit qubit coherence, providing an additional control parameter for quantum dot operation. Using these symmetric pulses we can tune the exchange interaction over two orders of magnitude, limited by the \AI{attainable voltage range and the lever arm of the plunger gates.} By optimising the gate stack to increase the lever arm of the plunger gates, the exchange tunability can be further enhanced in future devices. The currently available control still allows to implement a decoupled CZ two-qubit gate without pulsing on the barrier gate. These results \AI{enable the introduction of an architecture that does not rely on individual barrier gates, while crucially maintaining universal quantum control at a detuning-insensitive operating point. Consequently, the small barrier gates} of current semiconductor quantum processors \AI{can be merged into a grid in future designs}, facilitating \AI{scaling towards the sizes needed for useful }quantum simulation and computation.

\section*{Methods}

\subsection*{Initialisation and Readout}
\begin{figure}
    \centering
    \includegraphics[width=\linewidth]{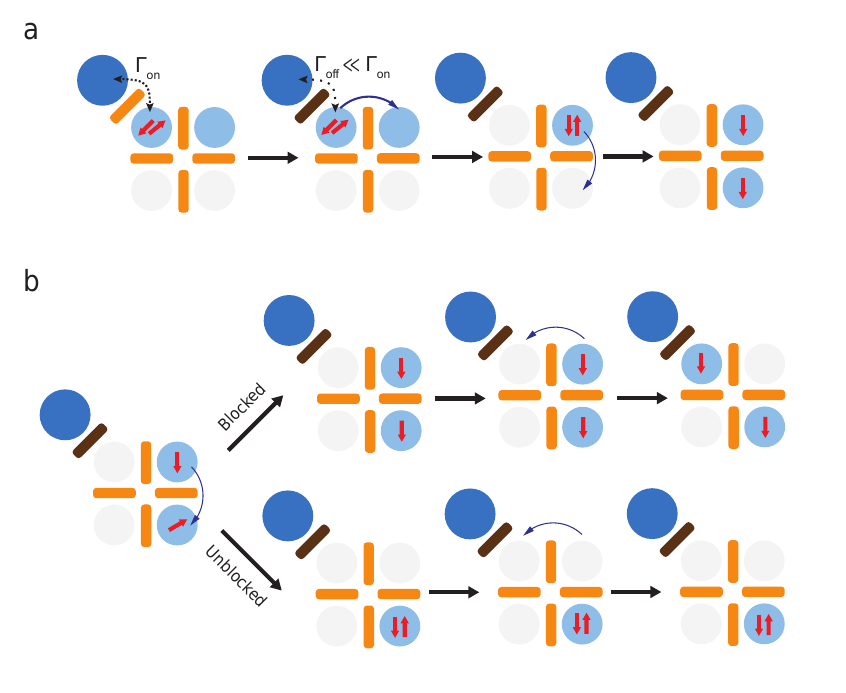}
    \caption{\textbf{Schematics of the qubit initialisation and readout} \textbf{(a)} For initialisation, spins are loaded from the reservoir and shuttled to the active quantum dots. After loading, the tunnel rate $\Gamma$ to the reservoir is decreased \textbf{(b)} For spin readout, after performing spin-to-charge conversion, any existing charge under plunger gate P\textsubscript{2} is shuttled to the charge sensor for readout.}
    \label{fig:InitReadout}
\end{figure}

At the beginning of each experimental shot, two hole spins are loaded onto quantum dot 1 from the reservoir. After loading, the tunnel coupling to the reservoir is decreased and the spins are shuttled from quantum dot 1 to quantum dot 2, where they are in a spin singlet. Through adiabatic transfer from the (0,2,0,0) to the (0,1,1,0) charge state, the $\ket{\downarrow \downarrow}$ spin state is formed in quantum dots 2 and 3. Figure~\ref{fig:InitReadout}a shows the schematic of this initialisation process. \AI{During the loading and initialisation, the voltages on virtual barrier gates $\text{\textmyfont{\textit{v}}}B_{23}$ and $\text{\textmyfont{\textit{v}}}B_{34}$ are pulsed by $\SI{-40}{mV}$ and $\SI{-50}{mV}$ respectively. Similar pulses are applied during the readout sequence described later. Considering the exchange tunability achieved with the barrier-free operation, we believe that symmetric barrier-free pulses can provide sufficient control over the tunnel coupling to initialise without using individual barrier gates. In particular we observe in Figure~\ref{fig:Exchange_data}e that $\mu$ can be used to realise the same exchange variation as an $\SI{125}{mV}$ pulse on the barrier gate. As a result, we foresee that in future work the entire qubit experiment can be performed without pulsing on the local barrier gate at all.}  \\

Spin readout starts with spin-to-charge conversion through Pauli spin blockade. This allows us to distinguish between singlet and triplet states. In our readout routine the $\ket{\downarrow \downarrow}$ basis state is mapped to the unblocked singlet, while the other states are blocked\AI{~\cite{kellyIdentifyingMitigatingErrors2025}}. Therefore to probe the dynamics of either spin state, we make sure the other spin state is $\ket{\downarrow}$. Following spin-to-charge conversion, we perform a charge shuttling step to enhance the charge signal during readout. After Pauli spin blockade between quantum dots 2 and 3, the unblocked spins lead to the (0,0,2,0) charge occupancy while the blocked charge state is still in the (0,1,1,0) occupancy. We lower the voltage on plunger gate 1 to allow any charges in quantum dot 2 to shuttle there. The (0,1,1,0) state becomes the (1,0,1,0), while the (0,0,2,0) is unchanged. This charge state difference is picked up by the charge sensors to determine the initial spin state. A schematic of this spin-to-charge conversion can be found in Figure~\ref{fig:InitReadout}b. \\

Each presented data point typically corresponds to $N_{\text{shots}}=300$ shots. The threshold to classify the sensor signal into blocked or unblocked states was put in the middle between the maximal and minimum values measured during these $N_{\text{shots}}$.

\section*{Acknowledgements}
We thank Sander de Snoo for continuous software development, and Olaf Benningshof and Chien-An Wang for maintenance and support of the experimental setup. We thank Francesco Borsoi for fruitful discussions.
\section*{Data Availability}
The code, analysis, and raw data supporting the findings of this study are openly available in a Zenodo repository: https://doi.org/10.5281/zenodo.15829323~\cite{zenodo_Data}.
\section*{Funding}
We acknowledge support by the Dutch Research Council through an NWO ENW grant and by the Dutch National Growth Fund (NGF), as part of the Quantum Delta NL programme. We acknowledge the European Union through ERC Starting Grant QUIST (850641) and through the IGNITE project of European Union’s Horizon Europe Framework Programme under grant agreement No. 101069515. This research was sponsored in part by the Army Research Office (ARO) under Award No. W911NF-23-1-0110. The views, conclusions, and recommendations contained in this document are those of the authors and are not necessarily endorsed nor should they be interpreted as representing the official policies, either expressed or implied, of the Army Research Office (ARO) or the U.S. Government. The
U.S. Government is authorized to reproduce and distribute reprints for Government purposes notwithstanding any copyright notation herein.

\section*{Competing Interest}
A.S. Ivlev is an inventor on a patent application on the merged barrier architecture
filed by Delft University of Technology under the application number NL38924. N.W.H. is a founder, and M.V., G.S. \AI{and L.V.} are founding advisors of Groove Quantum BV, with N.W.H., M.V., G.S. \AI{and L.V.} declaring equity interests. The remaining authors declare that they have no competing interests.


%

\end{document}


\onecolumngrid

\title{Supplementary Material: Operating spin qubits without individual barrier gates} 

\author{Alexander S. Ivlev}
\email{M.Veldhorst, m.veldhorst@tudelft.nl}
\affiliation{QuTech and Kavli Institute of Nanoscience, Delft University of Technology, PO Box 5046, 2600 GA Delft, The Netherlands}
\author{Damien R. Crielaard}
\affiliation{QuTech and Kavli Institute of Nanoscience, Delft University of Technology, PO Box 5046, 2600 GA Delft, The Netherlands}
\author{Marcel Meyer}
\affiliation{QuTech and Kavli Institute of Nanoscience, Delft University of Technology, PO Box 5046, 2600 GA Delft, The Netherlands}
\author{William I. L. Lawrie}
\affiliation{QuTech and Kavli Institute of Nanoscience, Delft University of Technology, PO Box 5046, 2600 GA Delft, The Netherlands}
\author{Nico W. Hendrickx}
\affiliation{QuTech and Kavli Institute of Nanoscience, Delft University of Technology, PO Box 5046, 2600 GA Delft, The Netherlands}
\author{Amir Sammak}
\affiliation{QuTech and Netherlands Organisation for Applied Scientific Research (TNO), Delft, The Netherlands}
\author{\AI{Yuta Matsumoto}}
\affiliation{QuTech and Kavli Institute of Nanoscience, Delft University of Technology, PO Box 5046, 2600 GA Delft, The Netherlands}
\author{\AI{Lieven M. K. Vandersypen}}
\affiliation{QuTech and Kavli Institute of Nanoscience, Delft University of Technology, PO Box 5046, 2600 GA Delft, The Netherlands}
\author{Giordano Scappucci}
\affiliation{QuTech and Kavli Institute of Nanoscience, Delft University of Technology, PO Box 5046, 2600 GA Delft, The Netherlands}
\author{Corentin Déprez}
\affiliation{QuTech and Kavli Institute of Nanoscience, Delft University of Technology, PO Box 5046, 2600 GA Delft, The Netherlands}

\author{Menno Veldhorst}
\affiliation{QuTech and Kavli Institute of Nanoscience, Delft University of Technology, PO Box 5046, 2600 GA Delft, The Netherlands}

\maketitle


\section{Simulation Of Symmetric Exchange Operation}
\begin{figure}
    \centering
    \includegraphics[width=0.6\linewidth]{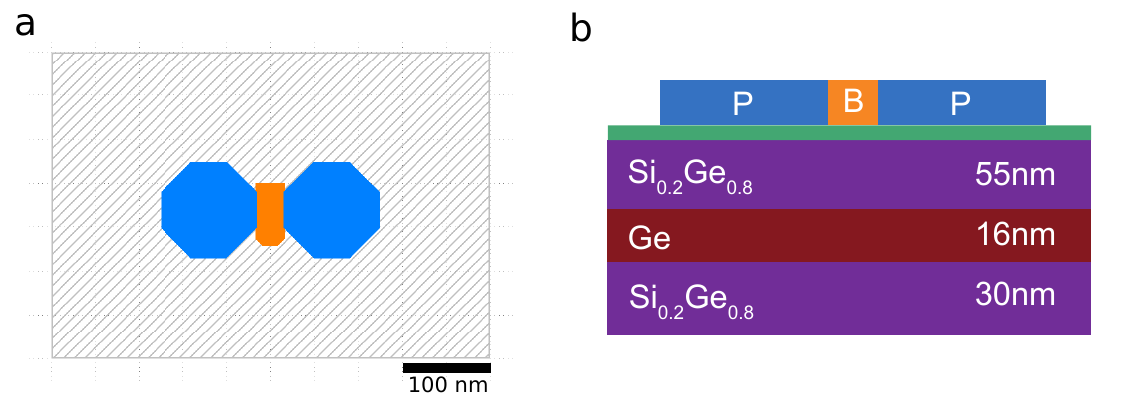}
    \caption{\AI{\textbf{Model of Simulated Device} \textbf{(a)} Top view of the model used in the simulation, with the two plunger gates in blue and the barrier gate in orange. The electric field and the wavefunction density is fixed at zero on the edge of the shaded area. \textbf{(b)} Schematic of the heterostructure used for the simulation. The gates are simulated as a 2D sheet and the green layer corresponds to the 1/10\,nm SiOx/AlOx dielectric.}}
    \label{fig:supp:SimulationModel}
\end{figure}

\AI{To verify experiment with theory we simulate a simplified model of the quantum device in QTCAD~\cite{QTCADAssistedDesign}, with the results included in Figure 1 of the main text. This model includes two plunger gates separated by a barrier gate, omitting the leads. The model and heterostructure used is detailed in Figure~\ref{fig:supp:SimulationModel}. The gates are simulated as a 2D perfectly conducting surface, with the plunger gate being defined in the same plane as the barrier gate. We note that this may explain why in the simulation the plunger gate pulse needs to be approximately double in voltage to achieve the same coupling as the barrier gate pulse, while in the experiment this factor was about three. The used heterostructure is similar to Figure~\ref{fig:supp:device}b but truncated $\SI{30}{nm}$ below the quantum well, as depicted in Figure~\ref{fig:supp:SimulationModel}b. A 1/10\,nm layer of SiOx/AlOx dielectric is included between the heterostructure and the gates.} \\

\AI{To compare the barrier- and plunger-based operation we apply various voltages on either the barrier or plunger gates, and solve the self-consistent Schr\"odinger-Poisson equation with the Luttinger-Kohn Hamiltonian truncated to 14 basis states. We extract the tunnel coupling as the difference between the ground state and the first excited orbital at zero detuning. We only consider tunnel coupling up to \SI{1}{meV} in order to limit the analysis to a well-defined barrier potential. In particular for the plunger-based operation, it is crucial to ensure that the system stays at the zero-detuning point. We minimise the energy splitting as a function of detuning to ensure this. While solving the Schr\"odinger equation, a tolerance of at most \SI{e-9}{eV} was used for plunger gate voltages below $\SI{-0.55}{V}$ and \SI{e-6}{eV} for the higher voltages. The simulation code can be found on Zenodo~\cite{zenodo_Data}, together with the other code used in the manuscript. }\\
\\

\AI{The fingerprint plot in Figure~1d of the main text is obtained by taking the analytical expression for the exchange energy $J=\frac{4U t_c^2}{U^2-\varepsilon^2}$ with a charging voltage of $U=\SI{35}{mV}$ and a lever arm of $\SI{0.09}{eV/V}$~\cite{john_2024_PRB_BichromaticRabiControl}, with an integration time window of $\SI{500}{ns}$. Detuning noise is included by averaging over a thousand independent realisations of $1/f$ charge noise with a lower cut-off frequency of $\SI{10}{mHz}$ and a noise amplitude $S_0=\SI{6.1E-10}{V^2/Hz}$~\cite{hendrickxSweetspotOperationGermanium2024,stehouwerExploitingEpitaxialStrained2025}.}

\section{Device and Fabrication}
\begin{figure}
    \centering
    \includegraphics[width=0.7\linewidth]{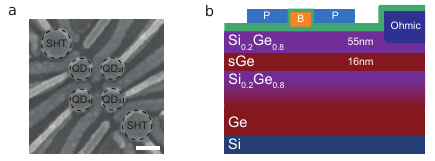}
    \caption{\AI{\textbf{Device Specifications} \textbf{(a)} An SEM of a nominally identical device to the one used in the experiment. A two-by-two array of quantum dots QD\textsubscript{1-4} is hosted under four plunger gates, with two single hole transistors at either site of the array. The scale bar corresponds to $\SI{100}{nm}$. \textbf{(b)} A schematic of the used heterostructure, with the relevant materials and thicknesses annotated. The green layer on top of the heterostructure and between the barrier (B) and plunger (P) gates corresponds to the SiOx/AlOx dielectric. The ohmic penetrates into the heterostructure as a result of thermal annealing.}}
    \label{fig:supp:device}
\end{figure}
\AI{The device is fabricated on a strained Ge/SiGe heterostructure (see Figure~\ref{fig:supp:device}). This heterostructure was grown through reduced-pressure chemical vapour deposition on a Si(001) wafer by reverse grading of $\text{Si}_{x}\text{Ge}_{1-x}$ from a concentration of $x=1$ to $x=0.2$. On top of the strain-relaxed Si\textsubscript{0.2}Ge\textsubscript{0.8} layer a strained $\SI{16}{nm}$ Ge quantum well is grown. Finally a $\SI{55}{nm}$ Si\textsubscript{0.2}Ge\textsubscript{0.8} buffer is grown and it is capped by a layer of native SiOx. This heterostructure has been shown to produce quantum mobilities as high as $\mu_q=\SI{2.5e4}{cm^2\; V^{-1}s^{-1}}$ and percolation densities of $p = \SI{2.1e10}{cm^{-2}}$~\cite{lodari_MatQuantTech_LowPercolationDensity2021}}. \\\\
\AI{The gate-structure is deposited in three metallic layers, all separated by $\SI{7}{nm}$ of AlOx grown through atomic layer deposition (ALD) at 300\textdegree C. The first metal layer consists of $\SI{30}{nm}$ thick aluminium, which during the subsequent ALD step diffuses down to the quantum well and provides an ohmic contact~\cite{hendrickx_naturecomm_GatecontrolledQuantumDots2018}. Subsequently Ti/Pd layers are deposited with thicknesses of 3/\SI{17}{nm} and 3/\SI{37}{nm} forming the barrier gates and plunger gates respectively. In this design the charge sensors are positioned diagonally with respect to the array, to mitigate any direct cross-talk that the inter-dot barrier gates have on the sensors. This device was first introduced in~\cite{hendrickx_2021_Nature_FourqubitGermaniumQuantum}.}
\section{EDSR Spectroscopy}
\begin{figure}[]
    \centering
    \includegraphics[width=0.8\linewidth]{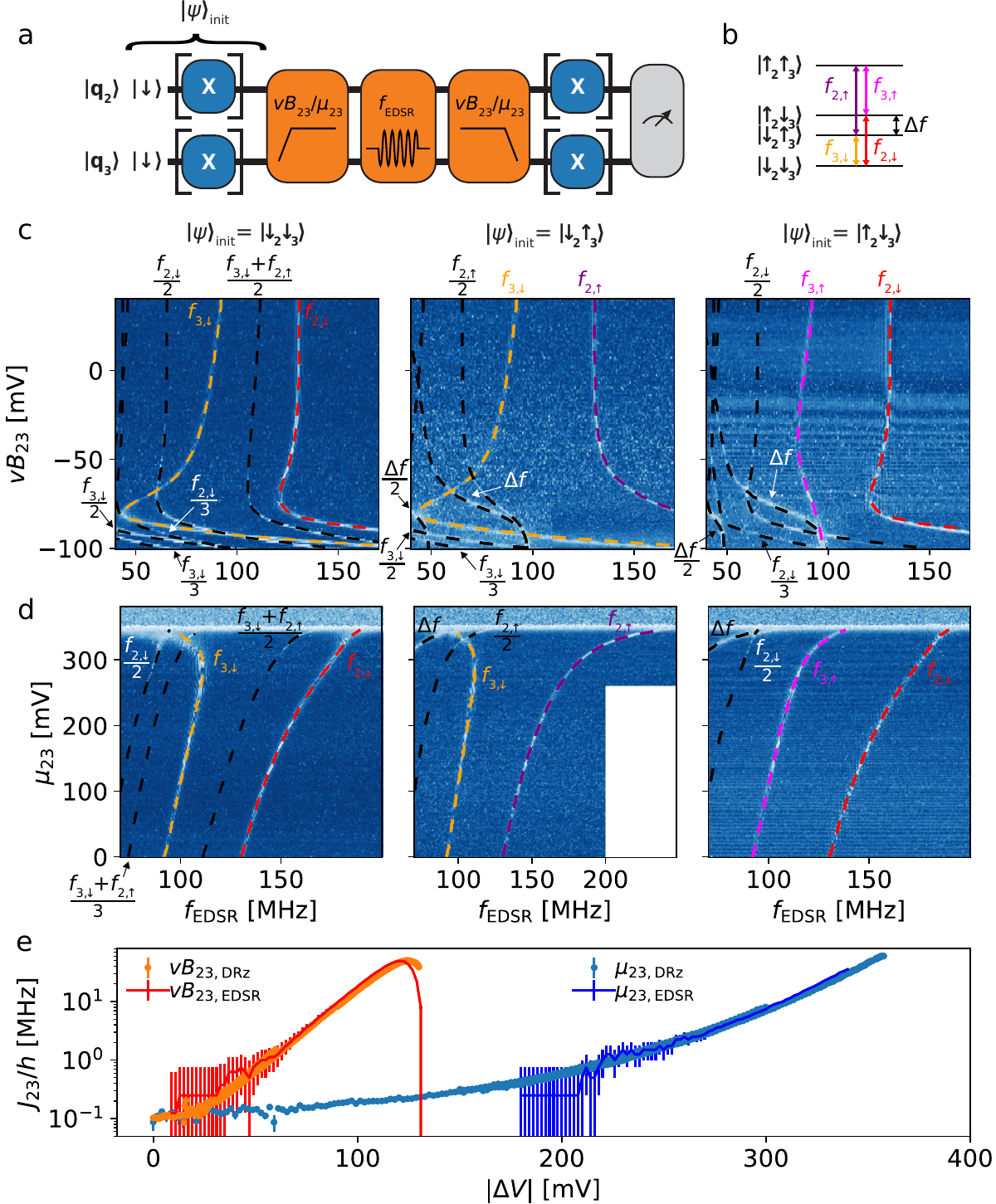}
    \caption{\textbf{Exchange splitting in EDSR spectra.} \textbf{(a)} The circuit used to perform the EDSR spectroscopy. Different spin states are initialised and their population is read out again after selectively applying X gates. In between, the EDSR frequency is swept at varying values for $vB_{\text{23}}$ or $\mu_{\text{23}}$. \textbf{(b)} Schematic annotating several transition frequencies. Subfigures \textbf{(c)} and \textbf{(d)} depict resonances detected through EDSR spectroscopy at different initial states $\ket{\psi}_{\text{init}}$, for different values of $vB_{\text{23}}$ and $\mu_{\text{23}}$ respectively. All observed resonances are annotated with the qubit transition frequencies or their harmonics. \AI{\textbf{(e)}} The exchange energy $J_{23,\text{EDSR}}$ is extracted from \textbf{c-d} and compared to the values from the main text, extracted using the decoupled CPhase (DRz) operation. The uncertainty in the EDSR value is set at $\SI{0.5}{MHz}$.}
    \label{fig:supp:EDSR_spec}
\end{figure}

We measure exchange splitting using EDSR spectroscopy to corroborate the exchange energy extracted from time-domain oscillations. The resonance frequencies are tracked for different initial states. The system starts in the $\ket{\downarrow_2 \downarrow_3}$, $\ket{\uparrow_2 \downarrow_3}$ or $\ket{\downarrow_2 \uparrow_3 }$ depending on whether respectively no qubit, $q_2$ or $q_3$ is flipped. After this state initialisation, the virtualised barrier voltage $vB_{23}$ or chemical potential $\mu_{23}$ is pulsed to the desired value and the resonances are probed by applying an EDSR pulse over a varying frequency $f_{\text{EDSR}}$ and fixed time $t_{\text{EDSR}} = \SI{1.6}{\micro s}$. The EDSR pulse is applied to plunger gate P\textsubscript{3}. After the EDSR pulse, the initialisation pulses are repeated to prepare for readout. The pulse sequence schematic can be found in Figure~\ref{fig:supp:EDSR_spec}a. The resulting resonance frequency maps for the different initial states as a function of $vB_{\text{23}}$ ($\mu_{23}$) can be found in Figure~\ref{fig:supp:EDSR_spec}c(d). The origin of all observed resonances is understood. The coloured resonances correspond to single spin-flip transitions (see the schematic in Figure~\ref{fig:supp:EDSR_spec}b), the values of which have been manually extracted from the spectroscopy data. The remaining resonances correspond to either subharmonics or double spin-flip transitions~\cite{john_2024_PRB_BichromaticRabiControl,saez-mollejoExchangeAnisotropiesMicrowavedriven2025}. We note that the artefacts at high values of $\mu_{23}$ correspond to the charge transition as also observed in the main text. Here, however, the onset of this transition is earlier due to the additional EDSR driving on plunger gate P\textsubscript{3}. \\

Having measured the spin-flip transition frequency from the EDSR data, we extract the exchange energy $J_{23}$ from it. Since $J_{23}$ equals the energy splitting of both spins depending on the state of the other spin, we take the average of these splittings for the value of the exchange. This results in $J_{23,\text{EDSR}}=\frac{f_{2,\uparrow}-f_{2,\downarrow}+f_{3,\uparrow}-f_{3,\downarrow}}{2/h}$, where $f_{i,\uparrow (\downarrow)}$ is the frequency of qubit $q_{i}$ conditioned on the other qubit being in state $\uparrow (\downarrow)$. In Figure~\ref{fig:supp:EDSR_spec}e we compare $J_{23,\text{EDSR}}$, with the exchange energy extracted using decoupled CPhase oscillations $J_{23,\text{DRz}}$ in the main text. Overall we see a strong correspondence between the two approaches, including the non-monotonicity. We see that the resolution and linewidth of the spectroscopy data do not allow us to measure exchange values below several hundred kilohertz. Also, some difference at large values of the exchange is observed, which can be attributed to AC Stark effects and linewidth broadening limiting the accuracy of the extracted resonance values. The uncertainty in the manually extracted frequencies is set at $\SI{0.5}{MHz}$. \\

\begin{figure}
    \centering    \includegraphics[width=0.8\linewidth]{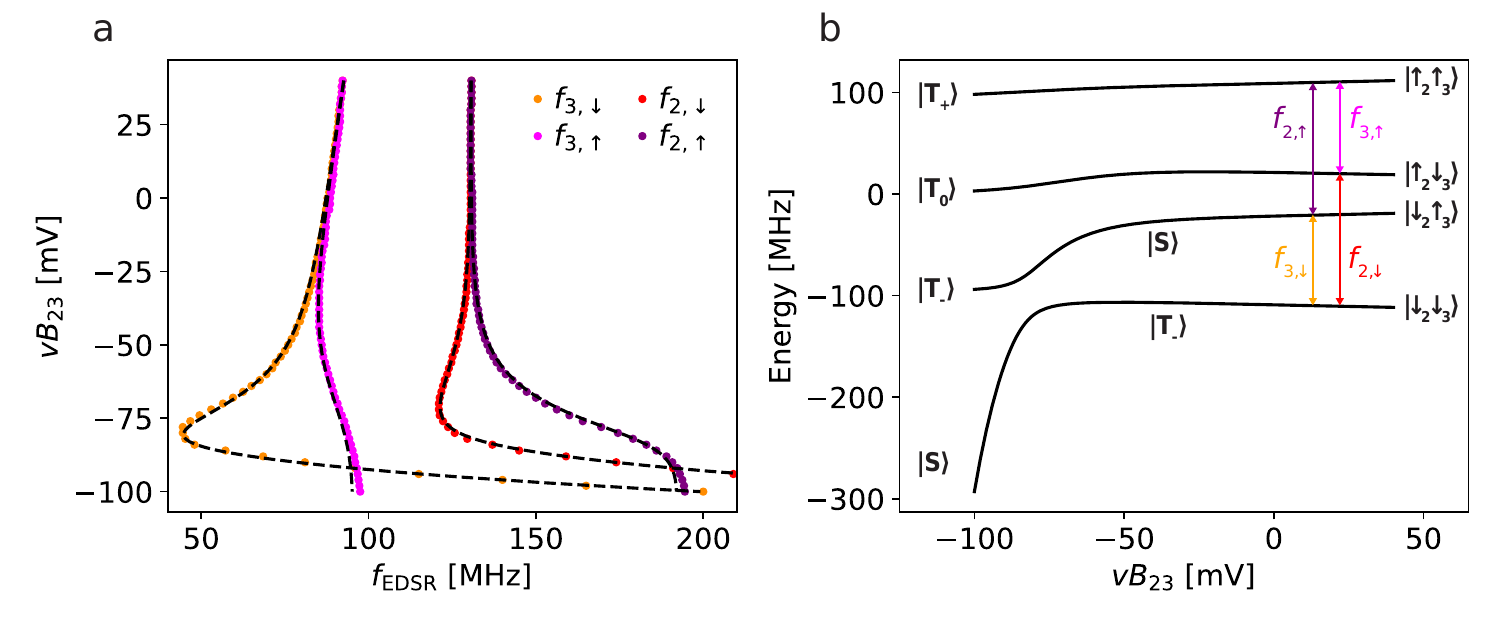}
    \caption{\textbf{Modeling EDSR spectrum}. \textbf{(a)} The EDSR spectrum data from Figure~\ref{fig:supp:EDSR_spec}c is fitted to the model of Equation~\ref{eq:model}. The resulting dashed line reproduces the data closely. \textbf{(b)} The energy eigenspectrum is plotted based on the parameters from the fit in subfigure \textbf{a}. The energy levels have been provided with labels, corresponding with the spin-states at high barrier voltage and the hybridised, singlet-triplet, states at lower barrier voltage.}
    \label{fig:supp:model}
\end{figure}

We note that at high $\mu_{23}$ and low $\text{\textmyfont{\textit{v}}}B_{23}$ the spin-states get hybridised and the resonance frequencies no longer correspond to a simple spin-flip. One consequence of this is the seemingly non-monotonic behaviour of $J_{23}$, which we hypothesise to result from the singlet $\ket{S}$ state anti-crossing with the triplet $\ket{T_-}$ state. To confirm this we simulate a four-level heavy hole system with the effective Hamiltonian~\cite{stepanenkoSinglettripletSplittingDouble2012,jirovecSinglettripletHoleSpin2021,saez-mollejoExchangeAnisotropiesMicrowavedriven2025} given by
\begin{equation}
    H = \begin{bmatrix}J(\text{\textmyfont{\textit{v}}}B_{23})&t_{\text{SO}}\sin\frac{\Omega}{2}&\frac{\Delta g \mu_B B_0}{2}\cos\frac
    {\Omega}{2}&t_{\text{SO}}\sin\frac
    {\Omega}{2}\\
    t_{\text{SO}}\sin\frac{\Omega}{2}&\frac{\sum g \mu_B B_0}{2}&0&0\\\frac{\Delta g \mu_B B_0}{2}\cos \frac{\Omega}{2}&0&0&0\\
    t_{\text{SO}}\sin\frac{\Omega}{2}&0&0&-\frac{\Sigma g \mu_B B_0}{2}
    \end{bmatrix}
    \label{eq:model}
\end{equation}
Here we assume $J(\text{\textmyfont{\textit{v}}}B_{23})=J_0\cdot 10^{\alpha\;\text{\textmyfont{\textit{v}}}B_{23}}$ varies exponentially with the barrier voltage. For $\sum g = g_2+g_3$ and $\Delta g = g_2-g_3$ we assume a linear dependence of the g-factor $g_{i} = g_{i,0}+g_{i,1}\text{\textmyfont{\textit{v}}}B_{23}$ on the barrier voltage. $B_0=\SI{50}{mT}$ is the applied magnetic field. We fit the EDSR spin-flip frequencies to this model, and with $$\{\Omega,t_{\text{SO}},g_{2,0},g_{3,0},g_{2,1},g_{3,1},J_{0},\alpha\}=\{0.636,\SI{75}{MHz},0.182,0.122,0,-2\cdot10^{-4},\SI{49}{kHz},\SI{26.9}{mV^{-1}}\}$$  we see a convincing agreement with between the model and the EDSR data, as depicted in Figure~\ref{fig:supp:model}. This validates our understanding that the apparent non-monotonicity in $J(\text{\textmyfont{\textit{v}}}B_{23})$ is a result from the $\ket
{S}-\ket{T_-}$ level crossing.

\section{Symmetric plunger operations on QD\textsubscript{1}-QD\textsubscript{4}}
\begin{figure}[H]
    \centering
    \includegraphics[width=0.7\linewidth]{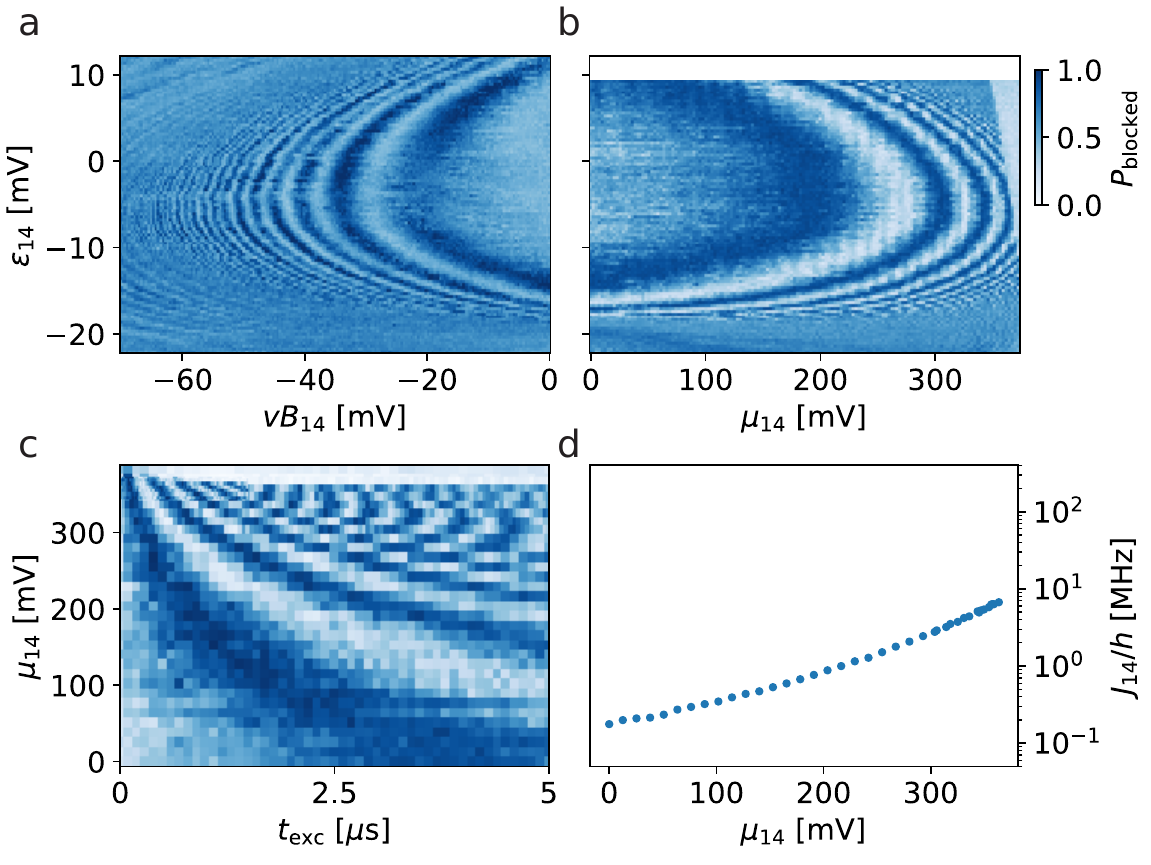}
    \caption{\textbf{Controlling exchange interaction with and without barrier gate pulses} \textbf{(a-b)} Fingerprint plots performed similar to the quantum dot pair in the main text, with the total exchange time is fixed at $t_{\text{exc}}=\SI{1}{\micro s}$. \textbf{(c)} Exchange oscillations as function of the exchange time $t_{\text{exc}}$ at the symmetry point $\epsilon_{14}\approx\SI{-5.5}{mV}$. \textbf{(d)} Exchange energy $J_{14}$ extracted from fitting the time-resolved oscillations in \textbf{c}.}
    \label{supp:fig:Pair41}
\end{figure}

The double quantum dot pair formed under plungers P\textsubscript{1} and P\textsubscript{4} have also been briefly analysed. \AI{To achieve the desired isolation, the north single hole transistor was depleted and instead the previously depleted south single hole transistor functioned as the charge sensor.} Detailed analysis and tuning were suspended due to \AI{a charge fluctuator, leading to a decrease in} quality of the south sensor signal. Still, we see the same qualitative behaviour on this double quantum dot as was observed in the main text for the QD\textsubscript{2}-QD\textsubscript{3} pair. Increasing $\mu_{14}\approx\frac{1}{2}(\text{\textmyfont{\textit{v}}}P_1+\text{\textmyfont{\textit{v}}}P_4)$ indeed enhances the exchange energy, and produces a fingerprint plot mirrored to the one for the virtual barrier voltage $\text{\textmyfont{\textit{v}}}B_{14}$, as seen in Figures~\ref{supp:fig:Pair41}a-b. This allows one to change the exchange coupling at fixed $\varepsilon_{14}\approx\frac{1}{2}(\text{\textmyfont{\textit{v}}}P_4-\text{\textmyfont{\textit{v}}}P_1)$. As observed with the previous pair, at large $\mu_{14}$ the charge is lost through a resonant level. We proceed to measure the exchange oscillations at varying $t_{\text{exc}}$ with a doubly decoupled sequence to fit the exchange energy (see Figures~\ref{supp:fig:Pair41}c-d), as is done in the main text. The time-resolved measurements are performed at $\varepsilon_{14}\approx\SI{-5.5}{mV}$, as this approximately correspond to the charge sweet spot. 

\section{Exchange Oscillation Quality}

\begin{figure}
    \centering
    \includegraphics[width=1\linewidth]{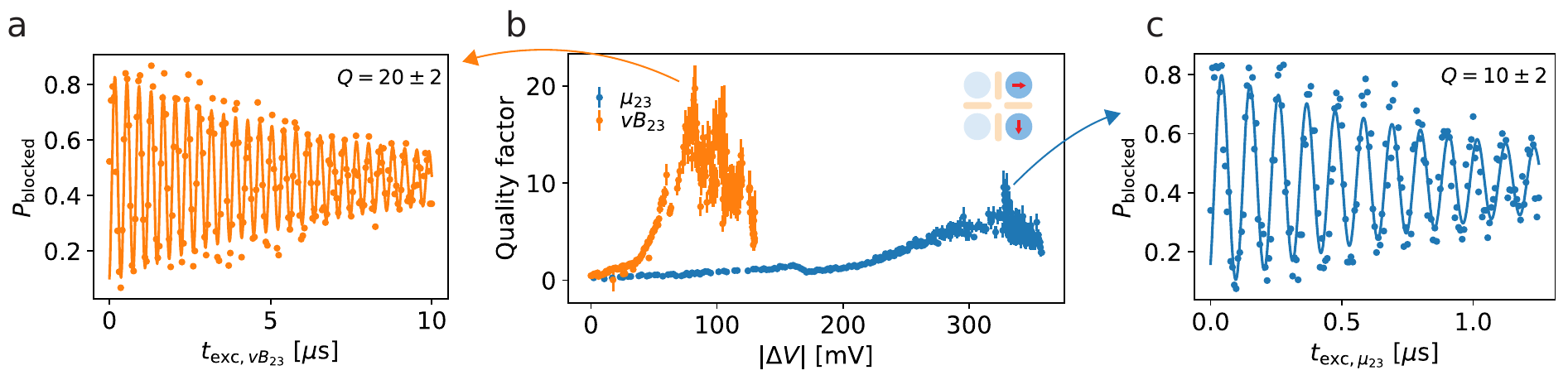}
    \caption{\AI{\textbf{Quality of $q_2-q_3$ exchange oscillations}. By extracting the characteristic decay time $T_{\text{decay}}$ and the frequency from the decaying sinusoidal fit, we determine the quality factor for a barrier-based and plunger-based exchange operation, for the qubits $q_2$ and $q_3$. The oscillations with the highest quality can be found in subfigures \textbf{a} and \textbf{c} for the barrier and plunger-based approach respectively, while subfigure \textbf{b} shows all the quality factors as function of the applied gate pulse amplitude. To allow for better fits, only characteristic decay times shorter than the measurement window were considered.}}
    \label{fig:supp:exchange_quality_23}
\end{figure}

\begin{figure}
    \centering
    \includegraphics[width=0.9\linewidth]{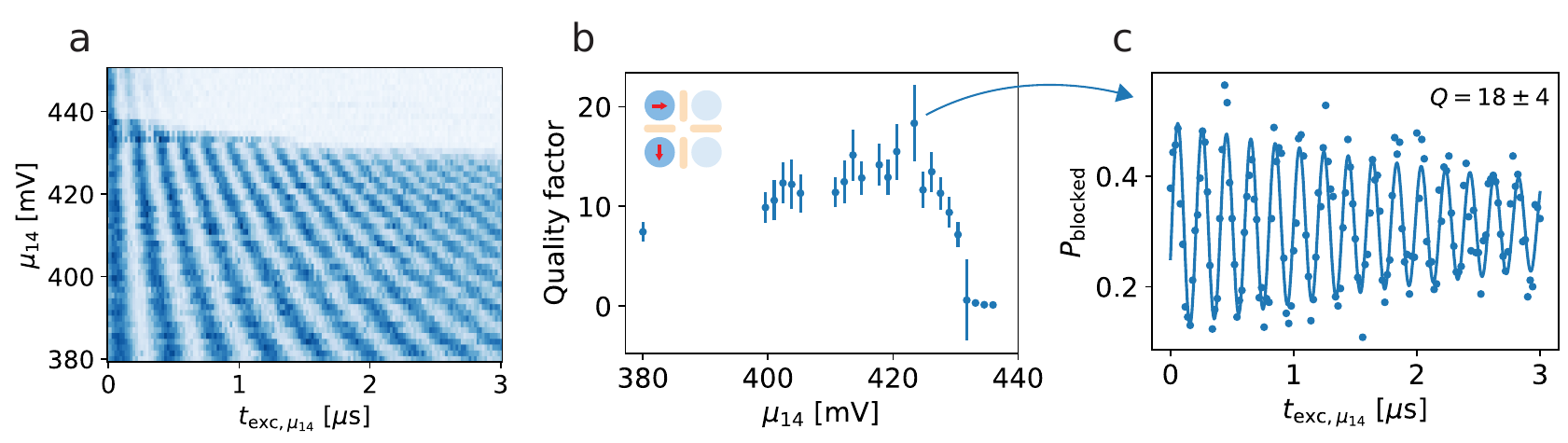}
    \caption{\AI{\textbf{Quality of $q_1-q_4$ exchange oscillations}. Similar to figure~\ref{fig:supp:exchange_quality_23} we extract the quality factor for qubits $q_1$ and $q_4$. The time-resolved oscillations are shown in subfigure \textbf{a}, noting that the DC-bias point is slightly different from the data shown in figure~\ref{supp:fig:Pair41} for this pair. Also, no data for the barrier-based approach is available here, as this pair was not investigated fully before the charge sensor signal degraded. Hence no faithful comparison can be made, but achievable quality can be gauged. The gap between the data-point at $\mu_{14}=\SI{380}{mV}$ and the other data-points results from filtering out the any decay times longer than compared 0.8 times the time-window of $\SI{3}{\micro s}$.}}
    \label{fig:supp:exchange_quality_14}
\end{figure}

\AI{To determine the performance of the symmetric barrier-free operation we extract the characteristic decay time $T_{\text{decay}}$ from the time-resolved exchange oscillations, by fitting them to $\sin(2\pi\cdot f_{\text{exc}}\cdot t_{\text{exc}}+\phi_0)\exp[-(t_{\text{exc}}/T_{\text{decay}})]$. Based on the fitted frequency $f_{\text{exc}}$ and the decay time we determine the quality factor $Q = T_{\text{decay}}\cdot f_{\text{exc}}$ of the oscillation. Figure~\ref{fig:supp:exchange_quality_23} shows these quality factors for different values of the virtual barrier voltage $\text{\textmyfont{\textit{v}}}B_{23}$ and symmetric plunger voltage $\mu_{23}$. The oscillations corresponding to the highest quality factor in each case have also been highlighted, respectively $Q_{\text{\textmyfont{\textit{v}}}B_{23}}=20\pm2$ and $Q_{\mu_{23}}=10\pm2$. Both approaches produce a lower quality factor than the typical values of $Q\approx20-50$ that is found in state-of-the-art devices in Si/SiGe heterostructures~\cite{weinsteinUniversalLogicEncoded2023,madzikOperatingTwoExchangeonly2025}. The difference may stem from the stronger spin-orbit interaction for holes in germanium, as well as variations in fabrication quality. Furthermore, we suspect that the noise-source limiting the coherence time of $q_3$ compared to $q_2$ (see Figure~2e of the main text) is also limiting the quality factor of the studied qubit pair. Initial data on the qubits under quantum dots QD\textsubscript{1}-QD\textsubscript{4} suggests that the symmetric plunger-based operation can produce higher quality factors up to $Q_{\mu_{14}}=18\pm4$ (see figure~\ref{fig:supp:exchange_quality_14}). To put the quality factor into perspective, using the fidelity estimate from \cite{stanoReviewPerformanceMetrics2022} a quality factor of $Q=18$ would result in a gate fidelity of $\mathcal{F}\approx 1-1/4Q\approx98.6\%$, assuming exponential dephasing. However assessing the performance of the barrier-free approach will require experiments like randomised benchmarking or gate-set tomography.} \\

%